\documentclass{appolb}
\usepackage{epsfig}
\usepackage{graphicx}

\begin{document}
\title{Sznajd model and its applications %
\thanks{Paper presented at First Polish Symposium on Econo- and Sociophysics}%
}
\author{Katarzyna Sznajd--Weron
\address{Institute of Theoretical Physics, University of
Wroc{\l}aw, pl. Maxa Borna 9, 50-204 Wroc{\l}aw, Poland }
}
\maketitle
\begin{abstract}
In 2000 we proposed a sociophysics model of opinion formation, which was based on trade union maxim "United we Stand, Divided we Fall" (USDF) and 
latter due to Dietrich Stauffer became known as the Sznajd model (SM). The main difference between SM compared to voter or Ising-type models is that information flows outward. 
In this paper we review the modifications and applications of SM that have been proposed in the literature. 
\end{abstract}
\PACS{05.50.+q Lattice theory and statistics (Ising, Potts, etc.) 89.65.-s Social systems}
  
\section{Introduction}
It was a typical winter morning in New York. Suddenly a man stopped and started to stare at the sky. 
People were passing by and almost nobody paid attention to this man. Next morning a group of four people 
started staring at the sky and $\ldots$ almost everybody walking by joined them. After several minutes 
the crowd blocked the street. What had happened? {\it Social validation} worked. The fundamental way that 
we decide what to do in a situation is to look to what others are doing \cite{Cialdini00}. We like to think 
about ourselves as individuals. Indeed, we are individuals but in many situations we behave like particles, 
which have no feelings and no free will. Not too many people are convinced by this statement but a number of experiments conducted by social psychologists confirm that \cite{Cialdini00,AWA94}. From a physicist's point of view the 
behavior of individuals and the interactions between them constitute a microscopic level of a social system. 
The question is if the laws on the microscopic scale of a social system can explain phenomena on the macroscopic scale, 
phenomena that sociologists deal with. 

Almost a century ago physicists asked the question if phase
transitions could be explained by microscopic theory. To answer
it, in 1920 Wilhelm Lenz proposed a very simple
microscopic model of interacting spins. Supervised by Lenz, Ernst
Ising in his dissertation (1924) studied the special case of a
linear chain of magnetic moments, which are able to take only two
positions - 'up' and 'down' - and which are coupled by interactions
between nearest neighbors. He showed that spontaneous
magnetization cannot be explained using this model in its
one-dimensional version \cite{Ising}. However, later it turned out that the two
dimensional version of the model (known presently as the Ising
model) can explain the critical phase transition. This taught us
that very simple local interactions can lead to qualitative
changes on the macroscopic scale.

Rapid changes on the macroscopic scale (like phase transitions in
physics) can be observed in various complex systems - from biological (e.g. mass
extinctions or speciation) to financial (crashes, speculative bubbles) or social
(sudden social depression or euphoria). These changes are usually
unexpected and no obvious source of such a behavior can be identified. In
recent years physicists have started to explain these "outside physics"
phenomena in terms of microscopic interactions, like they
have been doing for physical systems \cite{ladek1, ladek2}.

The Ising spin system is one of the most frequently used
models of statistical physics. 
Recently, this model has found numerous applications in other branches
of science including sociology \cite{g04,HKS01,fs05,MR03} and economy \cite{CS99,SMR99}. 
Of course, in sociophysics models the individual opinion can be described not only by Ising spins, but also by continuous variables
like in the model of Deffuant et al. \cite{DNAW00} or Krause and Hegselmann \cite{HK02}.

In 2000 we proposed a very simple model \cite{SWS00}, which was
aimed at describing global social phenomena (sociology) by local
social interactions (described by social psychology). We asked the
question on how opinions spread in a human society. Social opinion is
of course the outcome of individual opinions, represented in our model
by Ising spins ('yes' or 'no' ). Such an approach had been used
earlier \cite{g04,HKS01}.
However, motivated by the phenomenon known as {\it social validation}, we introduced 
a novel concept of spin dynamics \cite{SWS00}:
\begin{enumerate}
\item In each time step a pair of spins $S_i$ and $S_{i+1}$ is
chosen to change their nearest neighbors (nn), i.e. the spins 
$S_{i-1}$ and $S_{i+2}$.
\item If $S_i=S_{i+1}$ then
$S_{i-1}=S_{i}$ and $S_{i+2}=S_{i}$ ({\it social validation}).
\item If
$S_i=-S_{i+1}$ then $S_{i-1}=S_{i+1}$ and $S_{i+2}=S_{i}$. 
\end{enumerate}
In the model
two types of steady states are always reached - complete consensus (ferromagnetic state) or stalemate (antiferromagnetic state). The crucial difference of the model, originally called USDF after the trade union maxim "United we Stand, Divided we Fall", compared to voter or Ising-type
models is that information flows outward. The USDF model, later renamed by Dietrich Stauffer the "Sznajd model" (SM) has been modified  
and applied in marketing \cite{Schulze03,Schulze04,SWW03}, finance \cite{SWW02},
and politics \cite{stauffer02,nature,bcas01,BSK02,GSH04,Schneider04,SWS05,Stauffer03cm}; see also reviews \cite{fs05,Stauffer,Schechter}. Conversely, as has been noted by Slanina and Lavicka \cite{SL03}, sociologically inspired models pose new challenges to statistical physics. Therefore, SM has been investigated also from the theoretical point of view \cite{SL03,SO02,RSA04}.

\section{Modifications}

It has been claimed that the antiferromagnetic case can be considered to be quite unrealistic in a model trying to represent the behavior of a community. To avoid the unrealistic 50-50 alternating final state, new dynamic
rules were proposed in \cite{Sanchez04}:
\begin{enumerate}
\item In each time step a pair of spins $S_i$ and $S_{i+1}$ is
chosen to change their nearest neighbors (nn), i.e. the spins 
$S_{i-1}$ and $S_{i+2}$.
\item If $S_i=S_{i+1}$ then
$S_{i-1}=S_{i}$ and $S_{i+2}=S_{i}$ ({\it social validation}).
\item If
$S_i=-S_{i+1}$ then $S_{i-1}=S_{i}$ and $S_{i+2}=S_{i+1}$. 
\end{enumerate}

Other possibilities of avoiding the antiferromagnetic final state were briefly mentioned in \cite{SWS00}. These rules were called "if you do not know what to do, do nothing" and "if you do not know what to do, do whatever". 

Similarities and differences between SM and the well known voter model (VM) in one dimension were investigated in \cite{bs03}.
It was shown that the original one dimensional USDF rule is equivalent with a linear VM. Moreover, the model was expanded by considering synchronous updating rules. In the synchronous case, a asymmetric coexistence of the different opinions was also found in contrast to the original USDF model.
Synchronous updating was also investigated for two dimensional systems in \cite{sr03,sr04,a04,s04,YSKLL05} for several variants of SM. In particular, Stauffer \cite{s04} found  that in case of simultaneous updating a complete consensus is much more difficult. The reason is that for simultaneous updating some agents simultaneously receive conflicting messages from different neighbor pairs (called frustration) and thus refuse to change their opinion.

It is obvious that the one dimensional model is not very useful for social systems. Two dimensional models are much more realistic. Several possibilities of generalization to the square lattice were proposed by Stauffer et al. \cite{SSO00}. They presented the model on a square $L \times L$ lattice where again every spin can be up or down. Six different rules were introduced, but only two of them 
have been used in later publications:
\begin{itemize}
\item A $2 \times 2$ panel of four neighbors, if not all four center spins are parallel,
leaves its eight neighbors unchanged (see Fig.1).
\item A neighboring pair persuades its six neighbors to follow the pair orientation if
and only if the two pair spins are parallel.
\end{itemize}
With both these rules complete consensus is always reached as a steady state. Moreover, a phase transition is observed - initial densities below 1/2 of up-spins lead to all spins down and densities above 1/2 to all spins up for large enough systems \cite{SSO00}.

\begin{figure}
\includegraphics[scale=0.5]{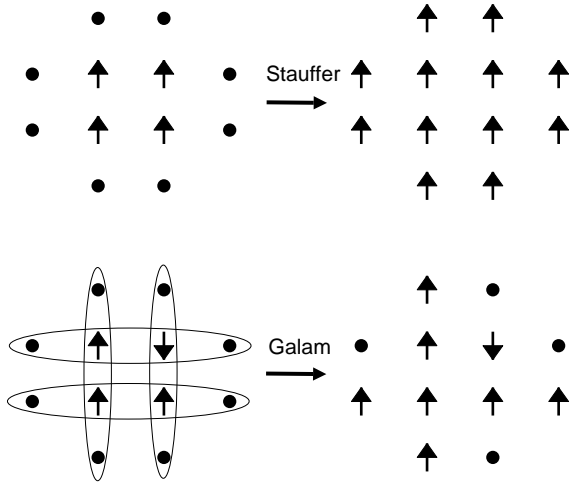}
\caption{Dynamical rules of SM on the square lattice proposed by Stauffer \cite{SSO00}(top) and Galam in private communication \cite{SSO00} (bottom) described in paper \cite{SSO00}. 
In every time step a panel of four spins denoted by arrows is picked at random and influences spins denoted by dots producing the configurations presented in the right-hand side pictures. Dots represent spins, which can be either up or down.}
\end{figure}

It seems that especially the first modification (presented in Fig.1), in which all four center spins have to be parallel to convince their neighbors, is very attractive from the social point of view \cite{AWA94}. To visualize this let us recall here some results of the classical Asch experiment on conformity.
Asch found that one of the situational factors that influence conformity is the size of the opposing majority. In a series of studies he varied the number of confederates who gave incorrect answers from 1 to 15. He found that the subjects conformed to a group of 4 as readily as they did to a larger group. However, the subjects conformed much less if they had an "ally". Apparently, it is difficult to be a minority of one but not so difficult to be part of a minority of two.
 
Galam (priv. comm. with Stauffer, described in \cite{SSO00}) showed that the updating rule of the one-dimensional SM can be transformed exactly into two dimensions in the following way (see Fig.1): the one-dimensional rule is applied to each of the four chains of four spins each, centered about two horizontal and two vertical pairs. His idea was used to construct two dimensional version of the so called two-component model \cite{SW04}. 

The generalization of SM to a triangular lattice with spreading of mixed opinion and with  pure antiferromagnetic opinion was studied in \cite{Chang01}, but no phase transition was found in this case. No phase transition is present also in the so called Ochrombel simplification \cite{Ochrombel01}. He proposed the model in which a randomly chosen spin on the square lattice influences its four neighbors, i.e. the neighbors get the same orientation. Thus, in contrast to the original model, we do not demand two or four people to share the same opinion to convince their neighbors. 

People do not rest on a site of a square lattice connected to their nearest neighbors. One step toward more realism would be a randomly diluted square lattice with many holes, which correspond to percolation. 
It was shown in \cite{MAS01} not much is changed if the square lattice is replaced by a
fractal structure coming from percolation, correlated or not - as in the regular square lattice, the system always found a fixed point and  again a phase transition near $p = 1/2$ appears. The
phase transition in SM seems very robust against changes in the geometry. 

The next step towards more realism is to investigate the model on a complex network (for review see \cite{AB01}). Up till now SM has been investigated on the small-world network \cite{E01,E03,HLL04}, on the scale-free networks \cite{BSK02,GSH04,b03,sousa05} and on the complete graph \cite{SL03}. For the latter case both simulation and analytical results were obtained \cite{SL03}. The evolution of SM with synchronous updating on several complex networks was also investigated \cite{YSKLL05}. Recently, unlike previous works, which ran SM dynamics on complex networks, this dynamics has been used to produce complex networks \cite{fc05}.

Changing topology of the chain to more complicated structure like two dimensional lattices or networks have certainly brought SM closer to reality. Another step toward reality has been done by increasing the range of interactions \cite{Schulze03a} and the number of variable's states \cite{stauffer02,SWS05,b03,s02a,fortunato04,Stauffer02IJMPC}, adding noise \cite{sr04,HLL04} or diffusion \cite{Schulze03,stauffer02}.

\section{Applications}
\subsection{Politics}
It seems that most successfully SM has been applied to politics. Modifications of SM allowed to reproduce the distribution of the number of candidates according to the number of votes they received in Brazilian and Indian elections \cite{bcas01,BSK02,GSH04}. 

Costa-Filho et al. \cite{caam99} showed that distributions of votes per candidate
for the 1998 elections in Brazil follow a power law distribution, with exponent
$\approx -1.0$. They obtained the same result for the whole country (candidates for a seat as federal deputy) as well as for the state of Sao Paulo (candidates for a seat as
state deputy). The same exponent was found for the elections of city councillors in Sao Paulo city in the elections of 2000. 
Based on SM, a model for proportional election with many-opinion modification
was proposed and good agreement with real political data was found \cite{bcas01}. 
The study showed that the empirically observed
 hyperbolic law has been a rather robust consequence of the modified SM, since the law was found first on the square
lattice \cite{bcas01} and then on both the cubic lattice and the Barabasi network \cite{BSK02}.

Another application of SM to politics is connected with the following question: 'What happens when there are several parties on the political stage, say two extremist and two centrist?' 
Such a situation pertained in the United Kingdom in the early 1980s, when the Liberals and the Social Democratic Party held the middle ground between the left-wing Labour Party and the right-wing Conservative Party. The two middle parties soon realized that there was not enough room for both of them, and merged in the late 1980s \cite{nature}.
The first approach to such a situation was made by Stauffer \cite{stauffer02,nature}.
In his modification of SM each lattice site initially is either empty, with probability 1/2, or has one of four possible opinions 1, 2, 3 or 4 (like in the Potts model), with probability 1/8 each. Then, at each time step every occupied site tries to move to an empty neighbor. Afterwards, randomly selected pairs of nearest neighbors, who share the same opinion, convince all those neighbors of the pair's opinion, which differ by at most one unit \cite{stauffer02}.
Stauffer found that parties 2 or 3 always win: they have more power of persuasion. But in most cases, one party other than the winner retained a small minority. This minority was always an extremist position, either 1 or 4.

Recently we have proposed another approach to describe political stage with four parties \cite{SWS05}. Our approach is based on the so-called Political Compass, which works by separating ideology into two major areas: economic and personal \cite{DN}. It allows us to discriminate between two kinds of behavior, connected with areas which we call personal and economic. It seems that an attitude with regard to the personal area can change in a different way than that with regard to the economic area. Thus, we assume that each
agent tries to influence its neighbors, but in the personal area
the information flows inward from the neighborhood (like in Glauber dynamics), whereas in the economic area the information flows outward to the neighborhood (like in SM).
Each person is characterized by two traits $(\sigma_i,S_i)$, where
$\sigma_i$ describes the attitude to personal freedom and $S_i$
describes the attitude to economic freedom. Both traits are
represented by Ising spins, i.e. binary variables (like in the
Ashkin-Teller model \cite{AT}). It turns out that independently of the initial attitudes
distribution (random or a single border) if we start with a group of agents who differ only in the
economic area the system always reaches an overall
consensus with all agents representing the same attitude.
If we start with a group of agents who differ only in the
personal area no consensus is possible if initially two attitudes are separated by a border line. However, in the case of random initial opinion distribution there is a chance for consensus or the system will divide into clusters. 

\subsection{Marketing}
How strong the advertising has to be in order to help one of the two products to win the whole market, even
though this product was initially in the minority \cite{Schulze03}? This question was asked in \cite{Schulze03,SWW03}. 
In both papers advertising is introduced as a kind of an external field, 
initial concentration of product A customers is $c_0$ and concentration of product B customers is $1 - c_0$. The
choice a customer makes (between two products sold in a duopoly market) is influenced
by the opinion of his/her neighbors and advertising. In our paper \cite{SWW03}, if the customer does not decide to follow the majority\'s
opinion he/she will be responsive to advertising. The response of the customer to
advertising(in this model only product A is being advertised) is measured by 
parameter $h$. There exists a
critical level of advertising $h$, as well as a critical value of initial concentration of customers above which product A will conquer 
the market with probability one (see Fig.2). Interestingly, a small level of advertising, such as 0.25, allows product A to win even if the initial concentration of A customers is very low $\approx 0.1$.
In the model proposed by Schulze \cite{Schulze03} advertising is included by assuming that at each iteration, every site becomes an A site with probability $h$. Moreover, he proposed to make the model more realistic by assuming that only half of the sites are occupied and agents 'diffuse' on the lattice. Additionally he assumed that feedback, which takes into account that advertising is diminished for an already successful product is possible. He simulated the model with and without feedback for two dimensional SM and the Ochrombel simplification. In agreement with our results, he found that a small fraction of advertising is sufficient to change nearly all
samples from product B to product A \cite{SWW03}. 

In real world oligopoly markets, innovation in terms of advertising and product
introduction has been critical in keeping the brand image contemporary. However, the
opportunity to create a sustainable competitive advantage through such actions is limited
in many markets because of the ease with which competitors can replicate the strategy. 
In our model \cite{SWW03} we found that above the critical value $h$ the relaxation of
the system is rather fast preventing an advertising counter-campaign. On the contrary,
if the level of advertising is close, but lower than $h$ the customers need much longer
times to make a final choice between the two products, which makes the advertising
campaign more costly and less efficient.

\begin{figure}
\includegraphics[scale=0.7]{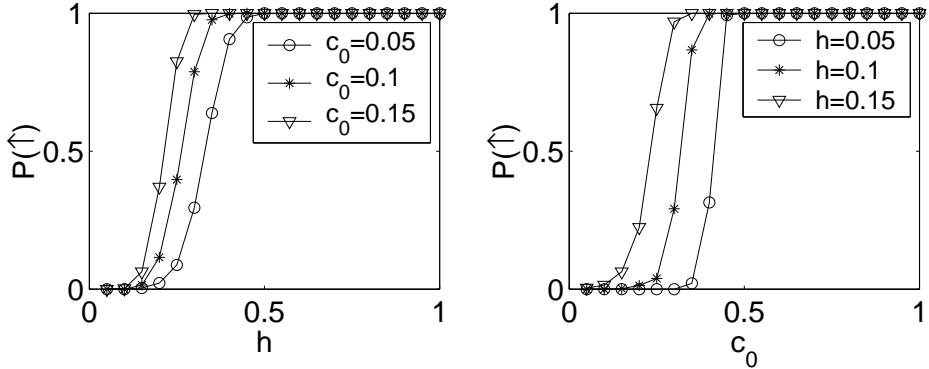}
\caption{\label{fig1} Probability P($\uparrow$) of reaching the final steady state with all up-spins as a function of the level of advertising $h$ (left panel) and the initial concentration $c_0$ (right panel).}
\end{figure}

\subsection{Finance}
Not much attention has been paid up till now to possible financial application of SM.
In this section we want to present the only application that has been proposed \cite{SWW02}.

In the model the spins are interpreted as market participants' attitude. An up-spin
($S_i = 1$) represents a trader who is bullish and places buy orders, whereas a down-spin
($S_i = -1$) represents a trader who is bearish and places sell orders. A similar approach was taken in \cite{CS99}.
In our model the first dynamic rule of the USDF model remains unchanged, i.e., if  $S_iS_{i+1} = 1$
then $S_{i-1}$ and $S_{i+2}$ take the direction of the pair $(i; i+1)$. This can be justified by the
fact that a lot of market participants are trend followers and place their orders on
the basis of a local guru's opinion. However, the second dynamic rule of the USDF
model has to be changed to incorporate the fact that the absence of a local guru
(two neighboring spins are in different directions) causes market participants to act
randomly. Of course, trend followers are not the only participants
of the market \cite{BPS97}. There are also rationalists - players that know much
more about the system and have a strategy. To make things simple, in our model we introduce one fundamentalist,
somewhat similar to Bornholdt's model \cite{Bornholdt01}. He knows exactly what is the current
difference between demand and supply in the whole system. If supply is greater
than demand he places buy orders, if lower - sell orders. 
The presented empirical analysis clearly shows \cite{SWW02} that three simple rules of our
model lead to a fat-tailed distribution of returns, long-term dependence in volatility
and no dependence in returns themselves as observed for market data (see Fig.3). Moreover, the obtained price trajectories possess characteristics
(returns of the price, lagged autocorrelation functions,
lagged autocorrelation functions of absolute value, Hurst exponent) in surprisingly good agreement with real data. 
Thus, it seems that this simple model is a good first approximation of a number of real
financial markets.

\begin{figure}
\includegraphics[scale=0.7]{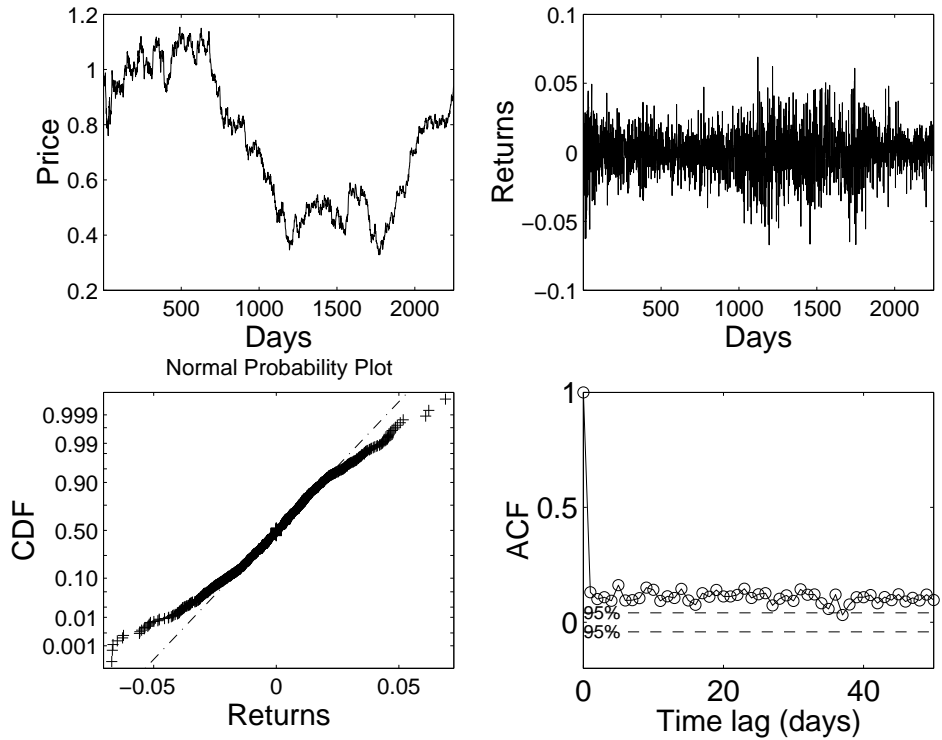}
\caption{\label{fig1} From the top left to the bottom right panel: a typical path of the simulated price process, returns of the simulated price process, normal probability plot of returns and lagged autocorrelation functions of absolute value of returns.}
\end{figure}

\section{Acknoledgements}
I am grateful to Prof. Ryszard Kutner, who encouraged me to write this paper.

\end{document}